\documentclass{amsart}

\usepackage{amssymb}
\RequirePackage{doi}
\usepackage{hyperref}
\usepackage{graphicx}
\usepackage[square,numbers]{natbib}

\usepackage[section]{placeins}
\usepackage{float}
\usepackage{comment}
\usepackage{xcolor}
\usepackage{graphicx}
\usepackage{dcolumn}
\usepackage{bm}

\newtheorem{theorem}{Theorem}[section]

\theoremstyle{definition}

\theoremstyle{remark}

\numberwithin{equation}{section}

\newmuskip\pFqmuskip
\newcommand*\pFq[6][8]{%
	\begingroup 
	\pFqmuskip=#1mu\relax
	\mathcode`\.=\string"8000
	\begingroup\lccode`\~=`\,
	\lowercase{\endgroup\let~}\pFqcomma
	{}_{\,#2}F_{\,#3}{\left[\genfrac..{0pt}{}{\,#4}{\,#5};#6\right]}%
	\endgroup
}
\newcommand{\pFqcomma}{\mskip\pFqmuskip}

\newcommand{\bbint}[2]{\ensuremath{\;\backslash\!\!\!\!\backslash\!\!\!\!\!\int_{#1}^{#2}}}

\begin{document}

\title[Finite-Part Integration]{Continuation of the Stieltjes Series to the Large Regime by Finite-part Integration}

\author{Christian D. Tica}
\author{Eric A. Galapon}
\address{Theoretical Physics Group, National Institute of Physics, University of the Philippines, Diliman Quezon City, 1101 Philippines}
\email{eagalapon@up.edu.ph}
\date{\today}

\begin{abstract}
        We devise a prescription to utilize a novel convergent expansion in the strong-asymptotic regime for the Stieltjes integral and its generalizations [Galapon E.A Proc.R.Soc A 473, 20160567(2017)] to sum the associated divergent series of Stieltjes across all asymptotic regimes. The novel expansion makes use of the divergent negative-power moments which we treated as Hadamard's finite part integrals. The result allowed us to compute the ground-state energy of the quartic, sextic anharmonic oscillators as well as the $\mathcal{PT}$ symmetric cubic oscillator, and the funnel potential across all perturbation regimes from a single expansion that is built from the divergent weak-coupling perturbation series and  incorporates the known leading-order strong-coupling behavior of the spectra. 
\end{abstract}

\date{\today}

\maketitle


\section{Introduction}
Divergent perturbation theory (PT) series is prevalent in many areas of theoretical physics \cite{le2012large}. They are ubiquitous in calculations of various physical quantities in
quantum field theory and in eigenvalue problems in quantum mechanics where exact and convergent solutions are rare. Hence a great deal of theoretical work is dedicated to the proper interpretation of these divergent solutions as well as procedures for extrapolating data in one parametric regimes to build the solution in the opposite regime \cite{olver1997asymptotics,alvarez2017new,costin2019resurgent}. A class of PT series solutions for which these problems have been addressed are the Stieltjes series,
\begin{equation}\label{mopy}
     F(\beta) =  \sum_{k=0}^{\infty}\mu_k(-\beta)^{k}, \qquad \beta\to 0,
\end{equation}
where $\mu_k$ are the positive-power moments of a positive function $\rho(x)$, $\mu_{k} = \int_{0}^{\infty} x^k \rho(x)\mathrm{d}x$.
The summation of a finite physical quantity $F(\beta)$ that is represented by a divergent series of Stieltjes can be carried out by various means when  $\beta$ is small \cite{bender1999advanced,gilewicz2002continued,weniger1990rational,weniger1993summation,mera2018fast,kazakov2002summation}. 
 Most notably, Pad\'{e} approximants constructed from a Stieltjes series possess well-understood convergence properties \cite{bender1999advanced, baker1996pade}.
 
 Determining $F(\beta)$ when $\beta$ is large, given only its leading-order behavior as $\beta\to\infty$ and a finite string of the positive-power moments $\mu_k$ is, by contrast, a formidable and long-standing problem in physics \cite{weniger1993summation,bender1994determination, suslov2001summing, le1990hydrogen}. In a turbulent transport process \cite{avellaneda1991integral}, the effective diffusivity $\kappa^*$, which exhibits $\kappa^{*}(Pe)\sim Pe^{1/2}$ as $Pe\to \infty$ is sought in the strong-P\'{e}clet number, $Pe$, regime given only the divergent, Stieltjes type small-P\'{e}clet number expansion. Similarly, in singular eigenvalue perturbation theory in quantum mechanics \cite{weniger1996construction, grecchiPadeSummabilityCubic2009, bender2001}, the spectrum $E(\beta)$ for the various anharmonic oscillators are computed in the strong coupling regime $\beta\to\infty$ given a finite number of the weak-coupling expansion coefficients and an algebraic power leading-order behavior in the strong-coupling regime $E(\beta)\sim\beta^{\lambda}$, $0<\lambda<1$. The Pad\'e approximants exhibit the leading-order behavior $P_M^N (\beta)\sim\beta^{N-M}$ as $\beta\to\infty$ where $N,M$ are integers, hence they will invariably fail to converge to $F(\beta)$ when $\beta$ is large in these cases \cite{weniger1996construction, dhatt2013embedding}.   

Some authors \cite{bender1987maximum,mead1984maximum} take a more direct route of summing a divergent Stieltjes series to a corresponding Stieltjes integral,
\begin{equation}\label{yib}
  \int_{0}^{\infty}\frac{\rho(x)}{1+\beta x}\mathrm{d}x\sim \sum_{k=0}^{\infty}\mu_k (-\beta)^{k}, \qquad \beta\to 0,
\end{equation}
by solving the underlying Stieltjes moment problem, that is the reconstruction of $\rho(x)$ from the available positive-power moments $\mu_{k}$ and obtaining an expansion suitable for computation even in the opposite asymptotic regime $\beta\to\infty$.  However, evaluating the Stieltjes integral in the strong asymptotic regime by expanding the geometric series $(1/(x\beta) + 1)^{-1}$ in powers of $1/(x\beta)$ and a term-wise integration is carried out leads to an expansion with terms diverging individually,
\begin{equation}\label{kir}
     \int_{0}^{\infty}\frac{\rho(x)}{1+\beta x}\mathrm{d}x \sim \sum_{k=0}^{\infty} \mu_{-(k+1)}\frac{(-1)^{k}}{\beta^{k+1}}
\end{equation}
 where the negative-power moments, $\mu_{-(k+1)}=\int_{0}^{\infty}\rho(x)/x^{k+1}\mathrm{d}x, k=0,1\dots$ are divergent integrals \cite{bender1994determination, weniger1996construction}. As a consequence, the evaluation of the Stieltjes integral is done numerically. In a seminal work \cite{galapon2}, one of us took this procedure further by giving these divergent negative-power moments $\mu_{-(k+1)}$ rigorous interpretations as complex contour integrals. This led us to derive a set of exact and convergent large-$\beta$ expansions for the Stieltjes integral as well as its generalizations \cite{tica2018finite, tica2019finite}.  

In this paper, we demonstrate a simple prescription for using these convergent expansions to sum the divergent Stieltjes series \eqref{mopy} for all values of the parameter $\beta$ especially in the formidable regime $\beta\to\infty$ where most summation technique fails. We discuss this procedure in Section \ref{meth} and apply it as a summation technique for divergent Stieltjes series and its generalization in Section \ref{appl}. To demonstrate the prescription, we will derive a strong-coupling expansion for the ground-state energy of the $\mathcal{PT}$ Symmetric cubic oscillator, the quartic anharmonic oscillator and the funnel potential from a finite collection of coefficients of the factorially diverging Rayleigh-Schrodinger(RS) weak-coupling perturbation expansions and their known non-integer algebraic power, leading-order behaviour in the strong-coupling regime. In addition, we shall also take on the more formidable RS weak-coupling expansion for the ground-state energy of the sextic anharmonic oscillator by summing it to a generalized Stieltjes integral. We then summarize and present several points to further extend and improve the efficacy of the method in Section \ref{summ}.
\begin{figure}
    \centering
	\includegraphics[scale=0.23]{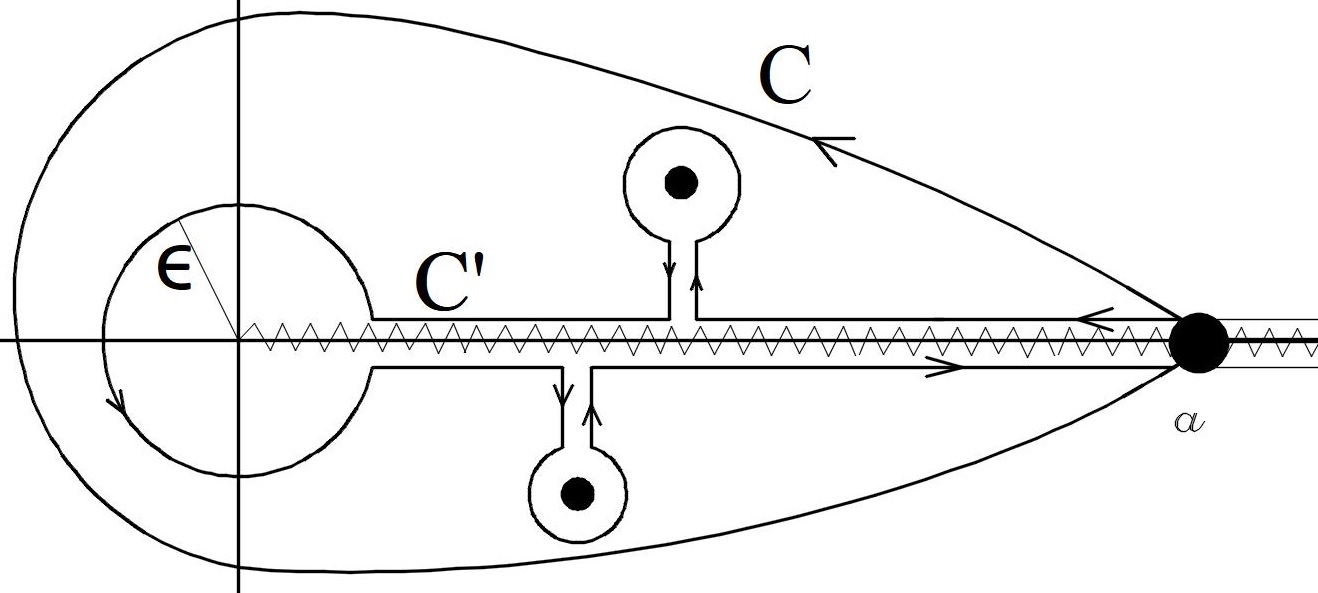}
	\caption{The contour of integration. The upper limit $a$ can be infinite. The points represent poles which for the case of the Stieltjes integral is at $z=-1/\beta$. $\epsilon$ is a small positive parameter.}
	\label{tear}
\end{figure}

\section{The method}\label{meth}
The prescription we propose allows us to circumvent the problem of divergent integrals arising in the formal procedure leading to equation \eqref{kir}. The solution lies in the interpretation of these divergent quantities as Hadamard's finite part \cite{monegato2009definitions, galapon2}. A concise discussion of the definitions and representations relevant in the present work is given in the Appendix \ref{hfp}. Because of the central role played by the finite parts for obtaining exact convergent expansions for the Stieltjes integral and its generalizations,  the method came to be known as finite-part integration. 

\subsection{Finite-Part Integration}
The problem of obtaining a convergent expansion in the $\beta\to\infty$ regime for the Stieltjes integral,
\begin{equation}\label{goy}
    S(\beta) = \beta\int_{0}^{\infty}\frac{\rho(x)}{1+\beta x}\mathrm{d}x,
\end{equation}
and its generalizations have been considered in a series of work \cite{ galapon2,tica2018finite,tica2019finite, galapon2016cauchy, villanueva2021finite, regularizedlimit}. A particular result which will be relevant in the application to be considered is the case when $\rho(x)= x^{-\nu}g(x)$ for $|\nu|<1$ and such that the complex extension $g(z)$ of $g(x)$ is entire. As in equation \eqref{kir}, the initial step is to deliberately induce divergent integrals by inserting the expansion for $(1/x\beta+1)^{-1}$ in powers of $1/x\beta$ and performing a formal term-by-term integration, 
\begin{equation}\label{panini}
\sum_{k=0}^{\infty}\frac{(-1)^{k}}{\beta^{k}}\int_{0}^{\infty}\frac{x^{-\nu}g(x)}{x^{k+1}}\mathrm{d}x.
\end{equation}
We then interpret the divergent integrals, the negative-power moments of $\rho(x)=x^{-\nu}g(x)$, as Hadamard's finite part integrals,
\begin{equation}\label{mp}
         \mu_{-(k+1)} = \int_{0}^{\infty}\frac{x^{-\nu} g(x)} {x^{k+1}}\mathrm{d}x \to      \bbint{0}{\infty}\frac{x^{-\nu} g(x)}{x^{k+1}}\mathrm{d}x.
\end{equation}
They are given rigorous representations as complex contour integrals in \cite{galapon2},
\begin{align}\label{result2}
\bbint{0}{a}\,\frac{x^{-\nu}g\left(x\right)}{x^{m}}\,\mathrm{d}x = \frac{1}{\left(e^{-2\,\pi\,\nu\,i}-1\right)}\,\int_{\mathcal{C}}\,\frac{z^{-\nu} g\left(z\right)}{z^{m}}\,\mathrm{d}z,
\end{align}
for $m=1,2,\dots$ and $a$ can be taken to infinity. The contour $\mathcal{C}$ is shown in Figure-\ref{tear}. 

The explicit calculation of the values of Hadamard's finite part integrals is described in the Appendix \ref{hfp} where we utilize their more expedient and familiar form \eqref{branchpoint} or as in \eqref{monir} where they are computed as analytic continuations of the Mellin transform. The main results of \cite{galapon2} and in the more recent work \cite{regularizedlimit} are to demonstrate the equivalence of these dual representations. The purpose of the contour integral representation \eqref{result2} of Hadamard's finite part integrals is to bring the evaluation of the Stieltjes integral \eqref{goy} into the complex plane using the equivalent contour $\mathcal{C'}$ as shown in Figure \ref{tear},
 \begin{align}
\int_{\mathcal{C'}}\frac{ z^{-\nu}\,g(z)}{1/\beta + z}\mathrm{d}z   = \left(e^{-2\pi\nu i}-1\right)    
\int_{0}^{a}\frac{x^{-\nu} g(x)}{1/\beta+ x}\mathrm{d}x 
+ 2\pi i \text{Res}\left[\frac{z^{-\nu}\,g(z)}{1/\beta + z}\right]_{z = -\frac{1}{\beta} }.
\end{align}
Solving for the Stieltjes integral along the real line, 
\begin{align}\label{boyt}
     \int_{0}^{a}\frac{x^{-\nu} g(x)}{1/\beta+ x}\mathrm{d}x = \frac{1}{\left(e^{-2\pi\nu i}-1\right)} \int_{\mathcal{C'}}\frac{ z^{-\nu}\,g(z)}{1/\beta + z}\mathrm{d}z 
     - \frac{2\pi i}{\left(e^{-2\pi\nu i} 
     - 1\right)}\,\text{Res}\left[\frac{z^{-\nu}\,g(z)}{1/\beta + z}\right]_{z = -1/\beta }.
\end{align}
We can now perform a term-by-term integration in the first term of the right-hand side of the equation \eqref{boyt},
\begin{align}\nonumber
        \frac{1}{\left(e^{-2\pi\nu i}-1\right)}  \int_{\mathcal{C'}}\frac{ z^{-\nu}\,g(z)}{1/\beta + z}\mathrm{d}z &= \sum_{k=0}^{\infty}\frac{(-1)^{k}}{\beta^k \left(e^{-2\pi\nu i} - 1 \right)}\int_{\mathcal{C'}}\frac{z^{-\nu}g(z)}{z^{k+1}}\mathrm{d}z\\
         & = \sum_{k=0}^{\infty}\frac{(-1)^{k}}{\beta^k} \bbint{0}{a} \frac{x^{-\nu} g(x)}{x^{k +1}}\mathrm{d}x,\qquad |z|>\frac{1}{\beta},
\end{align}
where we used the contour integral representation \eqref{result2}. Unlike the formal steps leading to equation \eqref{panini}, the expansion above can be shown to be absolutely convergent (see the proof of Theorem 3.4 in \cite{galapon2}). After evaluating the residue term in equation \eqref{boyt}, the Stieltjes integral \eqref{goy} admits the following exact and convergent expansion,
\begin{equation}\label{nion}
    \beta\int_{0}^{\infty}\frac{x^{-\nu} g(x)}{1+\beta x}\mathrm{d}x = \sum_{k=0}^{\infty}\frac{(-1)^{k}\mu_{-(k+1)}}{\beta^{k}} + \frac{\pi\,g(-\frac{1}{\beta})\,{\beta^{\nu}}}{\sin(\pi\nu)},
\end{equation}
where the negative-power moments $\mu_{-(k+1)}$ are the finite part integrals \eqref{mp}. The second term in expansion \eqref{nion}, which we recovered by integrating in the complex plane, is missed out by mere formal term-by-term integration as in equation \eqref{panini} followed by a regularization of the divergent integrals as Hadamard's finite part integrals. Provided that $g(0)\neq 0$, this missed-out term dominates the behavior of the Stieltjes integral as $\beta\to\infty$ \cite{tica2018finite, tica2019finite}.

We can also apply finite-part integration to some generalizations of the Stieltjes integral. For instance, performing exactly the same set of steps leads to the following exact convergent expansion,
    \begin{align}\label{piggy}
     \beta \int_{0}^{\infty} \frac{x^{-\nu} g(x)}{1+\beta x^{2}}\mathrm{d}x = \sum_{k=0}^{\infty} \frac{(-1)^{k}}{\beta^{k}}\mu_{-(2k+2)} + \Delta(\beta)
\end{align}
 where the residue term evaluates to
\begin{align}\label{tiy}
         \Delta(\beta) =  \frac{\pi\,\beta^{(1+\nu)/2}}{\sin{(\pi\nu)}} \left(\cos{\left( \frac{\pi\nu}{2}\right)\operatorname{Im}{g\left( i/\sqrt{\beta} \right)}}\right. 
         + \left.\sin{\left( \frac{\pi\nu}{2}\right)\operatorname{Re}{g\left(i/\sqrt{\beta} \right)}} \right),
\end{align}
assuming that $g(z)$ is real along the real line. The $\operatorname{Re}{z}$ and $\operatorname{Im}{z}$ are the real and imaginary parts of a complex number $z$, respectively.  The divergent negative power-moments $\mu_{-(2k+2)}$ are  again interpreted as Hadamard's finite part integrals
\begin{equation}
    \mu_{-(2k+2)} = \bbint{0}{\infty} \frac{x^{-\nu}g(x)}{x^{2k + 2} }\mathrm{d}x,\,\,k=0,1,\dots
\end{equation}
An important feature common to both expansions \eqref{nion} and \eqref{piggy} is that their second terms exhibit an algebraic leading-order behavior $\sim \beta^{\nu}$ and $\beta^{(1+\nu)/2}$, respectively, in the $\beta\to\infty$ regime. This implies that an \textit{a priori} knowledge on the behavior of these integrals as $\beta\to\infty$ may be used as a criterion for choosing an appropriate value for $\nu$ so that this leading-order algebraic behavior can be simulated by the second term. This will be of particular relevance in the physical examples that will be considered.

\section{Application to Summation Problems}\label{appl}
In this section, we discuss how finite-part integration can be applied to the Stieltjes summation problem: given a finite string of the positive-power moments $\mu_k$, that is, the expansion coefficients of the Stieltjes series,
\begin{equation}\label{biy}
     \int_{0}^{\infty}\frac{\rho(x)}{1+\beta x}\mathrm{d}x\sim \sum_{k=0}^{\infty}\mu_k (-\beta)^{k}, \qquad \beta\to 0,
\end{equation}
we wish now to obtain approximations to Stieltjes integral across all values of $\beta$ ideally in the $\beta\to\infty$ regime. 

In order to make use of the expansions \eqref{nion} and its generalization \eqref{piggy}, we need to reconstruct $\rho(x)$ given the first $d+1$ positive power moments $\mu_k$. We require that the reconstruction takes the form $\rho(x) = x^{-\nu} g(x)$ for some $|\nu|<1$ and $g(x)$ whose complex extension $g(z)$ is entire. This can be done by expanding $g(x)$ as a generalized Fourier series using the Laguerre polynomials as the orthonormal basis,
\begin{equation}\label{jits}
    g(x) = e^{-x/2}\sum_{m=0}^{\infty} c_m m! \sum_{k=0}^{m}\frac{(-x)^k}{(k!)^2 (m-k)!},
\end{equation}
subject to the condition that 
\begin{equation}\label{kot}
    \mu_k = \int_{0}^{\infty}x^{k-\nu}g(x)\mathrm{d}x, \qquad k=0,1,\dots, d.
\end{equation}
Substituting the reconstruction \eqref{jits} for $g(x)$ in equation \eqref{kot} leads to the system of $d+1$ linear equations for the first $d+1$ expansion coefficients $c_m$
\begin{equation}
    \mu_n = \sum_{m=0}^{d} c_m\, P(n,m)
\end{equation}
where the matrix $P(n,m)$ is given by 
\begin{equation}\label{viy}
    P(n,m) = m! 2^{n-\nu+1}\sum_{k=0}^{m}\frac{(-2)^{k}\Gamma(n+k-\nu+1)}{(k!)^{2} (m-k)!}.
\end{equation}

For the examples we considered in this paper, we solve the system \eqref{viy} by implementing a block algorithm for LU factorization \cite{strang1993introduction, demmel1997applied} with partial pivoting based on the LU function available in the C++ Boost uBLAS library. This allowed us to achieve parallel computation both at the application level using C++ Boost MPI library and logical threads using C++ Boost Thread library. We also used arbitrary-precision data types provided in Boost Multiprecision library to represent the positive-power moments $\mu_k$ in extended precision of up to three thousand digits and to perform our computations in arbitrary precision arithmetic to eliminate rounding errors. A good introduction on the C++ Boost libraries can be found on \cite{schaling2011boost}.  We also checked our results against the LU factorization provided by the Eigen 3 library \cite{eigenweb}. This library provides matrix and vector containers for arbitrary precision data type provided by the MPFR C++ library \cite{mpfr}. The extended-precision libraries we used all derive from GNU GMP \cite{gmp} and MPFR \cite{mpfr.org} libraries .

Once the reconstruction \eqref{jits} is obtained, the first term in \eqref{nion} can now be computed as 
\begin{equation}
    \sum_{k=0}^{\infty}\frac{(-1)^{k}\mu_{-(k+1)}}{\beta^{k}} = \sum_{k=0}^{\infty}\frac{(-1)^{k}}{\beta^{k}} \bbint{0}{\infty}\frac{x^{-\nu} g(x)}{x^{k+1}}\mathrm{d}x.
\end{equation}
Substituting the reconstruction \eqref{jits} and separating terms involving finite part integrals give
\begin{align}\nonumber
    \sum_{k=0}^{\infty}\frac{(-1)^{k}\mu_{-(k+1)}}{\beta^{k+1}} = \sum_{k=0}^{d} \frac{(-1)^{k}}{\beta^{k}} \left(A_k + B_k + C_k \right)
     + \sum_{k=d+1}^{\infty} \frac{(-1)^{k}}{\beta^{k}} D_k,
\end{align}
where
\begin{equation}\label{ah}
    A_k = \sum_{m=0}^{k}c_m m! \sum_{l=0}^{m}\frac{(-1)^{l}}{(l!)^2 (m-l)!}\bbint{0}{\infty}\frac{e^{-x/2}}{x^{k+\nu+1-l}}\mathrm{d}x,
\end{equation}
\begin{equation}\label{bah}
    B_k = \sum_{m=k+1}^{d}c_m m! \sum_{l=0}^{k}\frac{(-1)^{l}}{(l!)^2 (m-l)!}\bbint{0}{\infty}\frac{e^{-x/2}}{x^{k+\nu+1-l}}\mathrm{d}x,
\end{equation}
\begin{equation}\label{fop}
    C_k = \sum_{m=k+1}^{d}c_m m! \sum_{l=k+1}^{m}\frac{(-1)^{l}\,\Gamma(l-k-\nu) 2^{l-k-\nu}}{(l!)^2 (m-l)!},
\end{equation}
and
\begin{equation}\label{kah}
    D_k = \sum_{m=0}^{d}c_m m! \sum_{l=0}^{m}\frac{(-1)^{l}}{(l!)^2 (m-l)!}\bbint{0}{\infty}\frac{e^{-x/2}}{x^{k+\nu+1-l}}\mathrm{d}x.
\end{equation}
The finite part integrals appearing above are computed explicitly in the Appendix \ref{hfp} and is given by equation \eqref{expofinite},
\begin{align}\label{dah}
    \bbint{0}{\infty}\frac{e^{-x/2}}{x^{k+\nu+1-l}}\mathrm{d}x = \frac{(-1)^{k-l+1} \left(\frac{1}{2}\right)^{k-l+1+\nu}\pi}{\Gamma(k-l+1+\nu)\sin(\pi\nu)}.
\end{align}

The presence of the factor $g(-1/\beta)$ in the second term of the expansion \eqref{nion} reveals a crucial insight as to how the point-wise convergence of the reconstruction of $\rho(x) =x^{-\nu}g(x)$ near $x=0$ could undermine the convergence of $S(\beta)$ in regime $\beta\to\infty$. Of particular relevance to the reconstruction of $\rho(x) = x^{-\nu}g(x)$ as a generalized Fourier series \eqref{jits} is the Gibbs phenomenon \cite{arfken1999mathematical} about the origin. Spurious oscillation could result if the reconstruction of $\rho(x)$ could not properly simulate a zero or a singularity at $x=0$.  In this sense, the problem of summing the divergent Stieltjes series in this regime turns mainly into solving the underlying Stieltjes moment problem.
\subsection{Singular Perturbation Theory in Quantum Mechanics}
In the following subsections we will apply the procedure outlined above to the summation of divergent Rayleigh-Schrodinger (RS) perturbation series for the spectrum of Hamiltonian operators $\hat{H}(\beta) = \hat{H}_0 + \beta\hat{V}$, where $\beta$ is the coupling parameter that quantifies the degree by which the Hamiltonian $\hat{H}(\beta)$ departs from $H_0$ whose eigenfunctions and eigenvalues are known in closed form. The eigenvalue problems involving the Hamiltonians that we consider represent singular perturbation theory and the divergent perturbation theory (PT) series in powers of $\beta$ for the spectrum are interpreted as asymptotic expansions in the Poincar\'e sense \cite{simon2018tosio, simon1971anharmonic, loeffel1972cargese}. In each of these examples, a procedure of Symanzik scaling \cite{loeffel1972cargese} performs a unitary transformation on the Hamiltonian $\hat{H}(\beta)$ that turns the singular PT into a regular one and, more importantly, provides a strong-coupling leading-order behavior for the spectrum \cite{loeffel1972cargese, weniger1996construction}.

To demonstrate this procedure, consider the family of anharmonic oscillators, 
\begin{equation}\label{anhar}
    H(\alpha,\beta) = \hat{p}^{2} + \alpha\hat{x}^{2} + {\beta} x^{2m},\,\, m = 2, 3, 4,\dots
\end{equation}
and $\alpha ,\beta>0$. The RS weak-coupling expansion for the spectrum is,
\begin{equation}\label{pyr}
    E_n^{(m)}(\alpha,\beta) = \sum_{k=0}^{\infty}E_{k,n}^{(m)}(\alpha)\,\beta^{k}.
\end{equation}
Consider the unitary transformation,
\begin{equation}
    (U(\lambda) \psi)(x) = \lambda^{1/2}\psi(\lambda x), \,\, \lambda > 0,\,\,\,\,\psi(x) \in L^{2}(\mathbb{R}).
\end{equation}
So that,
\begin{equation}
    U(\lambda) \hat{x} U(\lambda)^{-1} = \lambda \hat{x},\qquad U(\lambda)\,\hat{p}\,U(\lambda)^{-1} = \lambda^{-1} \hat{p}.
\end{equation}
Applying the unitary transformation to \eqref{anhar},
\begin{equation}
        U(\lambda)H(\alpha,\beta)U(\lambda)^{-1} = \lambda^{-2} {H}(\alpha \lambda^{4}, \beta\lambda^{2m+2}).
\end{equation}
Hence the spectrum obey the scaling property
\begin{equation}
    E_{n}^{(m)}(\alpha,\beta) = \lambda^{-2} 
    {E}_{n}^{(m)}(\alpha \lambda^{4}, \beta \lambda^{2m+1}).
\end{equation}
For the case $\lambda = \beta^{-1/(2m+2)}$ and $\alpha=1$,
\begin{equation}
    H(1,\beta) = \hat{p}^{2}+\hat{x}^{2}+\beta\hat{x}^{2m}
\end{equation}
and $U(\lambda)H(1,\beta)U(\lambda)^{-1} = \mathcal{H}(\beta)$,
\begin{equation}\label{mak}
    \mathcal{H}(\beta) = \beta^{1/(2m+1)}(\hat{p}^{2} + \beta^{-2/(m+1)} \hat{x}^{2} + \hat{x}^{2m})
\end{equation}
so that the energy \eqref{pyr} of the $n$th state also possesses the strong-coupling expansion
\begin{equation}\label{bami}
   {E}^{(m)}_n (\beta) = \beta^{1/(m+1)}\sum_{k=0}^\infty K_{k, n}^{(m)}\beta^{-2k/(m+1)}
\end{equation}
which is convergent for sufficiently large $\beta$ since the unitarily equivalent Hamiltonian \eqref{mak} represents a regular perturbation theory (see the discussion of Kato-Rellich theorem on p. 384 of \cite{loeffel1972cargese}). 

The starting point of the prescription is to sum the weak-coupling expansion \eqref{pyr} to a Stieltjes integral \eqref{nion} for the quartic case $m=2$ and the generalized Stieltjes integral \eqref{piggy} for the sextic case $m=3$. We then utilize the leading-order strong-coupling behavior $E_n^{(m)}\sim\beta^{1/(m+1)}$ in the expansion \eqref{bami} as a criterion for choosing an appropriate value for the parameter $\nu$ in the expansions \eqref{nion} and \eqref{piggy} so that their dominant terms  could simulate this behavior in the strong-coupling regime. We also apply the prescription to two other systems whose weak-coupling expansion for the spectrum exhibit the same factorial divergence as that of the quartic anharmonic oscillator case.
\begin{table}
	\begin{tabular}{lll}
		\hline
		$\beta$  & Second term &  Third term, $\Delta(\beta)$ \\ 
		\hline
		$10^{-1}$ & $-1.8895969(10^{105})$ & $1.8895969(10^{105})$  \\
		$1.0$ & $-7.583247(10^{28}) $ & $7.583247(10^{28})$ \\
		$10$ & $3.56913(10^{5})$ & $-3.56913(10^{5})$ \\
		$10^{2}$ & $-1.0963506$ & $3.01245278$ \\
		$10^{4}$ & $-0.9991713$ & $7.2954004$ \\
		$10^{5}$ & $-0.99896648$ & $11.5613682$\\
		$10^{15}$ & $-0.99894396$ & $1156.1251$\\ 
		\hline
	\end{tabular}
	\caption{Comparison of terms in the expansion \eqref{gihk} for the  ground-state energy of the $\mathcal{PT}$  Symmetric cubic oscillator. Both terms are relevant in the weak to intermediate coupling regimes while in the strong-coupling regime, $\beta \to\infty$, the third term dominates the second term. }
	\label{boki}
\end{table}
\subsubsection{$\mathcal{PT}$  Symmetric Cubic Oscillator}
The RS weak-coupling expansion for the spectrum
\begin{equation}\label{buko}
    E_n(\beta) = E_n(0) + \sum_{k=1}^{\infty}e_{n,k}\beta^{k}, \qquad E_n(0) = 2n + 1
\end{equation}
of the $\mathcal{PT}$ symmetric 
non-Hermitian cubic oscillator
\begin{equation}\label{moy}
    \hat{H}\left(\beta\right) =  \hat{p}^{2}+ \hat{x}^{2}+i\sqrt{\beta} \hat{x}^{3},\qquad  \left(\beta>0\right)
\end{equation}
is divergent. This is evident in the factorial growth that the expansion coefficients $e_{n,k}$ exhibit
\cite{bender2001},
\begin{equation}
    e_{n,k} = \frac{-4\sqrt{15}}{(2\pi)^{3/2}}\left(-\frac{15}{8}\right)^{k}\Gamma\left(k+\frac{1}{2}\right)\left(1+\mathcal{O}\left(\frac{1}{k}\right)\right).
\end{equation}

The expansions \eqref{buko} have also been demonstrated to be once-subtracted Stieltjes series \cite{grecchiPadeSummabilityCubic2009,bender2001, grecchiSpectrumCubicOscillator2013}.  This implies, that $e_{n,k}$ are mapped to the positive-power moments of some positive function $\rho_n(x)$,
\begin{equation}
    e_{n,k+1} = (-1)^{k}\mu_{n,k} = (-1)^{k}\int_{0}^{\infty}t^{k}\rho_n(t)\mathrm{d}t,\,\, k =0,1,\dots
\end{equation}
so that the spectrum can be summed as a Stieltjes integral
\cite{grecchiSpectrumCubicOscillator2013, grecchiPadeSummabilityCubic2009}
\begin{equation}
    E_n(\beta) = E_n(0) + \beta\int_{0}^{\infty}\frac{\rho_n(t)}{1+\beta t}\mathrm{d}t.
\end{equation}
After the reconstruction of $\rho_n(t) = x^{-\nu} g(x)$ where $g(x)$ is the generalized Fourier series \eqref{jits}, we can then use the expansion of the Stieltjes integral by finite-part integration \eqref{nion} to evaluate the spectrum for all values of the perturbation parameter $\beta>0$. Furthermore,
Symanzik scaling $\hat{x}\to\beta^{-\frac{1}{10}}\hat{x}$ \cite{grecchiSpectrumCubicOscillator2013} transforms the Hamiltonian \eqref{moy} to 
\begin{equation}
    \mathcal{H}(\beta) =\beta^{1/5}\left(\hat{p}+\beta^{-2/5}\hat{x}^2+i\hat{x}^3\right).
\end{equation}
This provides a leading-order behavior $E_n(\beta)\sim\beta^{1/5}$ in the strong-coupling regime $\beta\to\infty$, which we mimic by setting $\nu = 1/5$ in equation \eqref{nion}. Hence, for the ground-state energy, we obtain the following convergent expansion
\begin{equation}\label{gihk}
    E_0(\beta) = 1 + \sum_{k=0}^{\infty}\frac{(-1)^{k}\mu_{-(k+1)}}{\beta^{k}} + \frac{\pi\,g(-\frac{1}{\beta})}{\sin(\pi/5)}\,\beta^{1/5}.
\end{equation}
 We tabulated the numerical values of the terms in the expansion \eqref{gihk} for various $\beta$ in Table \ref{boki}. It is clear that in the $\beta\to\infty$ regime, the third term which includes the factor $g(-1/\beta)$, which samples the reconstruction of $\rho(x) = x^{-\nu}g(x)$ near the origin, dominates the second term while remaining relevant as $\beta\to 0$. 
 
 The convergence of expansion \eqref{gihk} is summarized in Table \ref{bot}. For some values of the ground-state energy well into the strong-coupling regime where the method of Pad\'e approximant becomes unreliable, our result exhibits convergence in the first few digits. For small to intermediate values of $\beta$, we see an excellent agreement with the Pad\'e approximant $P_{1001}^{1000}(\beta)$ constructed from \eqref{buko} using two-thousand of the positive-power moments $\mu_{0,k}$ represented in up to 3000-digit precision. 
 
\begin{table}
\centering
	\begin{tabular}{  c l c l c l c l c l c   }
	
		\hline
		Moments & $\beta = 0.1 $ & $\beta = 0.2 $ & $\beta = 1.0 $ &  $\beta = 10 $ & $\beta = 100 $   \\ 
		\hline
		10 & $10^{-9}$  & $10^{-7}$ &  $10^{-5}$ &  $10^{-4}$ &  $10^{-3}$\\
		
		50 & $10^{-17}$ & $10^{-14}$ & $10^{-9}$ &  $10^{-4}$  & $10^{-3}$ \\
		
		100 & $10^{-24}$ & $10^{-18}$ & $10^{-11}$ &  $10^{-6}$ & $10^{-4}$ \\
		
		200 & $10^{-32}$ & $10^{-24}$ & $10^{-14}$ & $10^{-8}$ & $10^{-4}$  \\
		
		500 & $10^{-49}$ &  $10^{-36}$ & $10^{-20}$ &  $10^{-10}$ & $10^{-5}$ \\
		      
        1000 & $10^{-67}$ & $10^{-50}$&  $10^{-26}$ &  $10^{-12}$ & $10^{-6}$  \\ 
        
        1500 & $10^{-84}$ & $10^{-59}$ &  $10^{-30}$ &  $10^{-13}$ & $10^{-7}$  \\ 
        
        2000 & $10^{-95}$ & $10^{-69}$&  $10^{-34}$  & $10^{-15}$ & $10^{-7}$   \\ 
        		
	\hline
	\end{tabular}

	\begin{tabular}{ c l l l l}
	Moments used & $\beta = 10^3$ & $\beta = 10^4$ &  $\beta = 10^5$ &  $\beta = 10^6$\\ 
		\hline
        10 & \textcolor{blue}{4.}578585 & \textcolor{blue}{7.2}34330 & \textcolor{blue}{11.}4448 & \textcolor{blue}{18.}1181\\

        100 & \textcolor{blue}{4.60}4231 & \textcolor{blue}{7.2}89783 & \textcolor{blue}{11.5}481 & \textcolor{blue}{18.}2974 \\
        
		300 & \textcolor{blue}{4.60}5774 & \textcolor{blue}{7.29}4272 & \textcolor{blue}{11.5}575 & \textcolor{blue}{18.3}147\\
		
		500 & \textcolor{blue}{4.606}02 & \textcolor{blue}{7.29}517  & \textcolor{blue}{11.5}595  & \textcolor{blue}{18.3}187 \\ 
		
		1000 & \textcolor{blue}{4.606}16 & \textcolor{blue}{7.29}582  & \textcolor{blue}{11.56}12  & \textcolor{blue}{18.32}19 \\ 
		
		1500 & \textcolor{blue}{4.606}19 & \textcolor{blue}{7.296}03  & \textcolor{blue}{11.56}17  & \textcolor{blue}{18.32}30 \\ 
		
		2000 & \textcolor{blue}{4.6062}1  & \textcolor{blue}{7.296}12  & \textcolor{blue}{11.562}0  & \textcolor{blue}{18.32}36 \\
		\hline

		$P^{1000}_{1001}(\beta)$ &  4.35 592 & 4.35 592 & 5.11 080  & 5.11 080  \\

	\end{tabular}

	\begin{tabular}{cllll}
		\hline
		Moments used & $\beta = 10^{7}$ &  $\beta = 10^{15}$ & $\beta = 10^{16}$ & $\beta = 2(10^{20})$\\ 
		\hline
		10 & \textcolor{blue}{2}8.6946  & \textcolor{blue}{11}40.99 & \textcolor{blue}{18}08.32 & \textcolor{blue}{13}106.1\\

 		500 & \textcolor{blue}{29.0}313 & \textcolor{blue}{115}5.63 & \textcolor{blue}{183}1.54 & \textcolor{blue}{1327}4.6\\ 

		1000 & \textcolor{blue}{29.0}370 &
		\textcolor{blue}{115}5.90 &
		\textcolor{blue}{183}1.98 & \textcolor{blue}{1327}7.8\\

		2000 &  \textcolor{blue}{29.04}02  & \textcolor{blue}{1156.}06 & \textcolor{blue}{1832.}23 & \textcolor{blue}{1327}9.6\\ 

		\hline
		$P^{1000}_{1001}(\beta)$ & 5.12171 & 5.12182 & 5.12182 & 13.9605\\
		\hline
	\end{tabular}
    \caption{Convergence of the expansion \eqref{gihk} by finite-part integration for the ground-state energy of the $\mathcal{PT}$ symmetric cubic oscillator across all perturbation regimes. The order of magnitude of the error is computed against the Pad\'e approximant $P^{1000}_{1001}(\beta)$ which becomes unreliable as $\beta\to\infty$. }
	\label{bot}
\end{table}

\subsubsection{Quartic Anharmonic Oscillator}
The RS weak-coupling expansion for the spectrum of the quartic anharmonic oscillator
\begin{equation}
    \mathcal{H}(\beta) = \hat{p}^{2} + \hat{x}^{2} + \beta x^{4}
\end{equation}
are divergent for all $\beta>0$. For the ground-state energy, 
\begin{equation}\label{iby}
    E^{(2)}(\beta) = 1 + \sum_{k=1}^{\infty} b^{(k)}\beta^{k}.
\end{equation}
The coefficients of the expansion for the energy correction exhibit a factorial growth \cite{bender1971large},
\begin{equation}\label{moty}
    b^{(k)}\sim (-1)^{k+1}\frac{\sqrt{24}}{\pi^{3/2}}\,\Gamma\left(k+\frac{1}{2}\right)\left(\frac{3}{2}\right)^{k},
\end{equation}
and more importantly, the energy correction has been shown to be a series of Stieltjes and can thus be summed to a unique Stieltjes integral \cite{simonCouplingConstantAnalyticity1970}. This implies that the weak-coupling expansion coefficients $b^{(k)}$ are positive-power moments of some positive function $\rho(x)$,
\begin{equation}
   b^{(k + 1)} = (-1)^{k}\mu_{k} = (-1)^{k}\int_{0}^{\infty} x^{k} \rho(x)\mathrm{d}x
\end{equation}
for $k =0,1\dots$ . So that the ground-state energy admits a  Stieltjes integral representation,
\begin{equation}
    E^{(2)}(\beta) = 1 + \beta\int_{0}^{\infty}\frac{\rho(x)}{1+\beta x}\mathrm{d}x.
\end{equation}
    \begin{table}
	\begin{tabular}{c llllll}
		\hline
		Moments & $\beta = 10^2$ & $\beta = 10^3$ & $\beta = 10^7$ &  $\beta = 10^8$ & $\beta = 10^9$ & $\beta = 10^{19}$ \\ 
		\hline
		10 & \textcolor{blue}{4.9}56457 & \textcolor{blue}{10.}46254 & \textcolor{blue}{22}2.097 & \textcolor{blue}{4}78.313 & \textcolor{blue}{10}30.313 & \textcolor{blue}{22}19404 \\
		
		100 & \textcolor{blue}{4.99}7441 & \textcolor{blue}{10.6}1987 & \textcolor{blue}{22}7.111 & \textcolor{blue}{4}89.212 & \textcolor{blue}{10}53.891 & \textcolor{blue}{22}70382\\

 	500 & \textcolor{blue}{4.999}350 & \textcolor{blue}{10.63}710 & \textcolor{blue}{228}.008  & \textcolor{blue}{49}1.178  & \textcolor{blue}{10}58.161 & \textcolor{blue}{22}79646\\ 

		1000 & \textcolor{blue}{4.9994}09 & \textcolor{blue}{10.63}887 & \textcolor{blue}{228}.178 & \textcolor{blue}{491}.554 & 
		\textcolor{blue}{10}58.982 & \textcolor{blue}{228}1434 \\ 

		2000 & \textcolor{blue}{4.99941}69 & \textcolor{blue}{10.639}53 & \textcolor{blue}{228}.283 & \textcolor{blue}{49}1.790 & \textcolor{blue}{10}59.498 & \textcolor{blue}{228}2561\\ 
		\hline
		$\mathcal{E}(\beta)$ & 4.9994175& 10.63979 & 228.450 & 492.177 & 1060.362 & 2284481 \\
        \hline
        $P^{1000}_{1001} (\beta)$ & 4.9991097 & 9.999928 & 13.9598 & 13.9604 & 13.9605 & 13.9605\\

        \hline
	\end{tabular}
	\begin{tabular}{  c l c l c l c l c    }
		Moments used & $\beta = 0.1 $ & $\beta = 0.2 $ & $\beta = 1.0 $ & $\beta = 4 $ & $\beta = 10 $   \\ 
		\hline
		10 & $10^{-9}$  & $10^{-7}$ &  $10^{-4}$ & $10^{-4}$ & $10^{-3}$  \\
		
		500 & $10^{-48}$ &  $10^{-28}$ & $10^{-19}$ & $10^{-12}$ & $10^{-9}$   \\
		    
        1000 & $10^{-66}$ & $10^{-48}$ &  $10^{-25}$ & $10^{-15}$ & $10^{-11}$  \\ 
        
        1500 &  $10^{-80}$ & $10^{-60}$ & $10^{-29}$  & $0^{-17}$ & $10^{-13}$  \\ 
        
        2000 & $10^{-92}$ & $10^{-67}$ & $10^{-33}$  & $10^{-19}$ & $10^{-14}$  \\ 
        
        3000 & $10^{-115}$  & $10^{-83}$& $10^{-38}$  & $10^{-22}$ & $10^{-16}$  \\ 		
	\hline
	\end{tabular}
    \caption{Convergence of the exapansion \eqref{ito} for the ground-state energy of the quartic anharmonic oscillator across all coupling regimes. The order of magnitude of the error is computed against the Pad\'e approximant $P^{1000}_{1001} (\beta)$ for $\beta = 0.1 - 10$ and $\mathcal{E}(\beta)$ is obtained using the result in \cite{weniger1996construction}.}
	\label{honggo}
\end{table}
In addition, the ground-state energy also exhibits a leading-order strong-coupling behavior by Symanzik scaling in equation \eqref{bami}, $E(\beta)\sim \beta^{1/3}$. Hence, applying the expansion \eqref{nion} and setting  $\nu=1/3$ to incorporate this behavior in the second term yields the convergent expansion 
\begin{align}\label{ito}
    E^{(2)}(\beta) = 1 + \sum_{k=0}^{\infty}\frac{(-1)^{k}\mu_{-(k+1)}}{\beta^{k}} + \frac{\pi\,g(-\frac{1}{\beta})\,\beta^{1/3}}{\sin(\pi/3)}.
\end{align}

 The result for computing $E^{(2)}(\beta)$
 across all coupling regimes is summarized in Table \ref{honggo}. We reproduced the first few digits of the result derived from the strong-coupling expansion \eqref{bami} which is obtained in \cite{weniger1996construction} using a significantly more complex algorithm. For various small and intermediate values of $\beta$,  we computed the error of our result against the Pad\'e approximant $P^{1000}_{1001} (\beta)$. For values of $\beta$ up to $\sim 10^{2}$, we see an excellent agreement between the result of our prescription and that of the Pad\'e approximant. For stronger couplings, the latter becomes unreliable.

\subsubsection{Funnel Potential}
The Hamiltonian with a linear confining potential,
\begin{equation}
    H(\beta) = \frac{p^{2}}{2} - \frac{1}{r} + \beta r, \qquad p = -i \mathbf{\nabla},
\end{equation}
    is relevant in quantum chromodynamics to model heavy quarkonium states \cite{eichten1978charmonium}. The spectrum can be obtained perturbatively using the method of logarithmic perturbation \cite{eletsky1981logarithmic}. The perturbation expansion for the ground-state energy in particular, 
    \begin{equation}\label{bonta}
        E(\beta) = -\frac{1}{2}\sum_{k=0}^{\infty}\epsilon_k (-\beta)^k, \qquad \epsilon_0 = 1
    \end{equation}
    diverges for all values of the perturbing parameter $\beta>0$ as evident in the factorial growth,
\begin{equation}
        \epsilon_k\sim \frac{36}{\pi e ^{3}} k! \left(\frac{3}{2}\right)^{k} k, \qquad k\to\infty.
    \end{equation}
    \begin{table*}
	\begin{tabular}{ c l l l l c c c}
		\hline
		Moments & $\beta = 1$ & $\beta = 10 $ &  $\beta = 18$ \\ 
		\hline
         10 & $\textcolor{blue}{0.577}313$ & \textcolor{blue}{6.}053 & 9.7803 \\
         
        100 & $\textcolor{blue}{0.5779213}441$ & $\textcolor{blue}{6.14}241$ & 9.99991 
        \\
		200 & $ \textcolor{blue}{0.5779213519}479$  & $\textcolor{blue}{6.143 }333  $ & $\textcolor{blue}{10.00}49751$ 
         \\
		500 & $ \textcolor{blue}{0.5779213519615935}03$ & $ \textcolor{blue}{6.14343}32$ & $ \textcolor{blue}{10.0059}2966$ 
        \\
		1000 & $ \textcolor{blue}{0.5779213519615935659650}094$ & $ \textcolor{blue}{6.143434}59 $ & $\textcolor{blue}{10.00596}453$
        \\
	   1250 & \textcolor{blue}{0.57792135196159356596508}592 & $\textcolor{blue}{6.1434346068}578$  & $\textcolor{blue}{10.0059653}2$ \\

		\hline
		$E^{*}(\beta)$ & $0.577 921 35$ & $6.143 4346$ & $10.005967$  \\
  
		\hline
		$P^{624}_{625} (\beta)$ & $0.57792135196159356596508625$ & $6.1434346088273$ & $10.00596539$ \\
  
		\hline
	\end{tabular}

 	\begin{tabular}{ c l l c c c}
		Moments &  $\beta=62.5$ &  $\beta=100$ &  $\beta=10^3$ & $\beta=10^6$  & $\beta=10^{11}$\\ 
		\hline
        10 & \textcolor{blue}{2}3.833 &  $\textcolor{blue}{3}3.20$ & ${1}60.6324$ & ${1}6264.15$ & 35044767.8\\
        
        100 & $\textcolor{blue}{24.}77139$ &  $\textcolor{blue}{34.}71469$  & 171.2058 & 17487.35 & 37684845.6  \\
        
		200 & $\textcolor{blue}{24.8}2973$ & $ \textcolor{blue}{34.}83 425 $ & $172.7231$ & $17706.17$ & 38158567.2\\
  
		500 & $\textcolor{blue}{24.85}288 $ & $ \textcolor{blue}{34}.8919 $ &  $173.9565$ & $17926.99$ & 38638242.3\\
  
		1000  & $\textcolor{blue}{24.85}589$ & $ \textcolor{blue}{34.90}229$ & $174.4599$  & $18053.81$ & 38915266.1\\
  
        1250 & $\textcolor{blue}{24.856}117$ &  $\textcolor{blue}{34.90}335$ & $174.5643$& $18088.61$ & $38991664.5$\\

		\hline

        $E^{*}(\beta)$ & $24.856630$ & $34.90444$ & - & - & - \\
        \hline
        
		$P^{624}_{625} (\beta)$ & $24.854283$ & $34.88143$ & $148.0630$ & $286.9456$ & $289.84609$ \\
		\hline
        
	\end{tabular}
	
	\caption{Convergence of the expansion \eqref{kwi} by finite part integration of the ground-state energy for the funnel potential across all coupling regimes. $E^{*}(\beta)$ are quoted from \cite{ganster1986modified, arteca1984method} while $P^{624}_{625} (\beta)$ is a Pad\'e approximant constructed from the PT series \eqref{bonta} for the energy correction.}
	\label{holat}
\end{table*}
    of the coefficients $\epsilon_k$. We sum the expansion as a once-subtracted Stieltjes series by mapping the coefficients as positive-power moments 
    \begin{equation}
        \epsilon_{k+1} =\mu_k = \int_{0}^{\infty} x^{k} \rho(x) \mathrm{d}x, \qquad k =0,1,\dots
    \end{equation}
Hence, the ground-state energy assumes a once-subtracted Stieltjes integral representation \cite{eletsky1981logarithmic}
\begin{align}
        E(\beta)\nonumber
         = -\frac{1}{2}\left(1 - \beta\sum_{k=0}^{\infty}\mu_{k} (-\beta)^{k} \right) = -\frac{1}{2}\left(1 - \beta \int_{0}^{\infty} \frac{ \rho(x)}{1+\beta x} \,\mathrm{d}x \right). 
\end{align}
Symanzik scaling $r\to\beta^{-1/3} r$ yields $H(\beta) = \beta^{2/3} H\left(\beta^{-1/3}\right)$ so that the ground-state energy exhibits a leading-order behavior $E(\beta) \sim \beta^{2/3}$ in the strong-coupling regime. By setting $\nu=2/3$ so that the reconstruction takes the form $\rho(x) = x^{-2/3} g(x)$ with $g(x)$ given by equation \eqref{jits}, we incorporate this leading-order behavior in the expansion \eqref{nion} of the Stieltjes integral so that we obtain the strong-coupling expansion,
\begin{align}\label{kwi}
    E(\beta) = -\frac{1}{2}\left(1 - \sum_{k=0}^{\infty} \frac{(-1)^k \mu_{-(k+1)}}{\beta^k} - \frac{\pi g(-\frac{1}{\beta})\beta^{2/3} }{ \sin\left(2\pi/3\right) } \right).
\end{align}

We tabulated the result of computing the ground-state energy $E(\beta)$ using the expansion \eqref{kwi} across all coupling regimes in Table \ref{holat}. We also quoted the results from \cite{ganster1986modified, arteca1984method} and the Pad\'e approximant $P^{1000}_{1001} (\beta)$ for comparison. As in the previous systems, our results are in excellent agreement with those derived from other methods in the weak-coupling regimes. Furthermore, armed with mostly the same information as that in Pad\'e approximants, we were also able to access a wider set of parametric regimes especially the strong-coupling domain where the Pad\'e approximant breaks down.   

\subsubsection{Sextic Anharmonic Oscillator}
Unlike the previous systems, the coefficients $b_3 ^{(k)}$ of the divergent weak-coupling expansion
\begin{equation}\label{tip}
    E^{(3)}(\beta) = 1 + \sum_{k=1}^{\infty} b_{3}^{(k)}\beta^{k}
\end{equation}
for the correction to the ground-state energy of the sextic anharmonic oscillator
\begin{equation}
    {H}(\beta) = \hat{p}^{2} + \hat{x}^{2} + \beta x^{6},\qquad \beta>0.
\end{equation}
exhibits a more pathological rate of divergence \cite{weniger1993summation, weniger1996construction},
\begin{equation}\label{giyit}
        b_{3}^{(k)} \sim (-1)^{k+1}\frac{\sqrt{128}}{\pi^{2}}\Gamma\left(2k+\frac{1}{2}\right)\left(\frac{4}{\pi}\right)^{2k},\,\, r\to \infty.
\end{equation}
In Table \ref{hoj}, we tabulated the result for approximating the ground-state energy from a sequence of partial sums of the divergent RS weak-coupling expansion \eqref{tip}. This table shows that due to the rate by which the expansion \eqref{tip} diverges, summation procedure is necessary even in the weak-coupling regime. 
\begin{table}
	\begin{tabular}{ clll }
		\hline
		Terms used  & $\beta = 10^{-3}$ & $\beta = 10^{-2}$ &  $\beta = 10^{-1}$\\ 
		\hline
		1 & $ \textcolor{blue}{1.0018}750$  & $\textcolor{blue}{1.01}875 $ & $ \textcolor{blue}{1.1}87500 $ \\
		5 & $ \textcolor{blue}{1.00184881}776  $ & $ \textcolor{blue}{1.01}7593128 $ & $ 1.302(10^{2}) $\\
		10 & $ \textcolor{blue}{1.0018488155}4160 $ & $ 3.56(10^{-1}) $ & $ -7.57(10^{9})$\\
		20 & $ \textcolor{blue}{1.00184881}39131  $ & $ -2.22(10^{11}) $ & $ $ \\
		\hline
		$E^{(3)}(\beta)$ & $1.00184881557231$ & $1.016741136375$ & $1.109087078466$ \\
		\hline
	\end{tabular}
	\caption{Convergence of the partial sums of the divergent weak-coupling PT expansion \eqref{tip} for the ground-state energy of the sextic anharmonic oscillator. $E^{(3)}(\beta)$ is obtained from our result in expansion \eqref{bxi}.}
	\label{hoj}
\end{table}
As remarked in \cite{weniger1993summation}, the growth rate \eqref{giyit} can be simulated by a Poincar\'e-type asymptotic expansion of a generalized Stieltjes integral
\begin{equation}
    \int_{0}^{\infty}\frac{t^{-1/2}e^{-t}\mathrm{d}t}{1+(64\beta t^{2}/45\pi^{2})}\sim \pi^{1/2}\sum_{k=0}^{\infty}\left(\frac{1}{2}\right)_{2k}\left(\frac{-64\beta}{45\pi^2}\right)^{k}.
\end{equation}
Hence, we sum the weak-coupling expansion \eqref{tip} for the energy correction to a generalized Stieltjes integral \eqref{piggy} by mapping the RS expansion coefficients $b_{3}^{(k)}$ to the $\mu_{2k}$ positive power moments of some positive function $\rho(x)$,
\begin{equation}\label{bos}
        b_{3}^{(k+1)} = (-1)^{k}\mu_{2k} = (-1)^{k}\int_{0}^{\infty} x^{2k} \rho(x)\mathrm{d}x,\qquad k=0,1,\dots
\end{equation}
 so that the energy correction in equation \eqref{tip} is summed as 
\begin{align}\nonumber
    \beta \sum_{k=0}^{\infty} b_{3}^{(k+1)}\beta^{k} & = \beta \sum_{k=0}^{\infty} (-1)^{k}\beta^{k}\int_{0}^{\infty}x^{2k}\rho(x)\mathrm{d}x\\ \label{kita} \nonumber
    &=\beta \int_{0}^{\infty}\rho(x)\left(\sum_{k=0}^{\infty} (-1)^{k}\beta^k x^{2k}\right) \mathrm{d}x \\
    &=  \beta \int_{0}^{\infty} \frac{\rho(x)}{1+\beta x^{2}}\mathrm{d}x.
\end{align}

Hence, the ground-state energy admits a generalized Stieltjes integral representation
\begin{equation}\label{podti}
    E^{(3)}(\beta) = 1 + \beta \int_{0}^{\infty} \frac{\rho(x)}{1+\beta x^{2}}\mathrm{d}x,\qquad \beta>0.
\end{equation}
We then set $\rho(x) = x^{-\nu} g(x)$, $|\nu|<1$ and $g(0)\neq 0$ so that the ground-state energy possesses the following strong-coupling expansion from the result in equation \eqref{piggy},
\begin{equation}\label{bxi}
    E^{(3)}(\beta) = 1 + \sum_{k=0}^{\infty} \frac{(-1)^{k}}{\beta^{k}}\mu_{-(2k+2)} + \Delta(\beta)
\end{equation}
where the missed term is given by equation \eqref{tiy} ,
\begin{align}\label{ugag}
         \Delta(\beta) =  \frac{\pi\,\beta^{(1+\nu)/2}}{\sin{(\pi\nu)}} \left(\cos{\left( \frac{\pi\nu}{2}\right)\operatorname{Im}{g\left( i/\sqrt{\beta} \right)}}\right. 
         + \left.\sin{\left( \frac{\pi\nu}{2}\right)\operatorname{Re}{g\left(i/\sqrt{\beta} \right)}} \right),
\end{align}
and the divergent negative-power moments $\mu_{-(2k+2)}$ are the Hadamard's finite part integrals
\begin{equation}\label{tuy}
    \mu_{-(2k+2)} = \bbint{0}{\infty} \frac{x^{-\nu}g(x)}{x^{2k + 2} }\mathrm{d}x,\,\,k=0,1,\dots
\end{equation}
From equation \eqref{bami}, the ground-state energy exhibit a leading-order behavior $E^{(3)}\sim \beta^{1/4}$  which we can incorporate in the term \eqref{ugag} by setting $\nu = -1/2$. As in equation \eqref{jits}, the function $\rho(x) = x^{1/2} g(x) $, is reconstructed as a generalized Fouries series expansion in terms of the Laguerre polynomials,
\begin{equation}\label{ibt}
    \rho(x) = x^{1/2}\,e^{-x/2}\sum_{m=0}^{\infty} c_m m! \sum_{k=0}^{m}\frac{(-x)^k}{(k!)^2 (m-k)!}.
\end{equation}
In the present case, the first $d+1$ expansion coefficients $c_m$ are computed by imposing equation \eqref{bos}. This leads to a system of linear equations for the expansion coefficients $c_m$, given the first $d+1$ coefficients $b_3 ^{(k+1)}$,
\begin{equation}
    b_3 ^{(k+1)} = \sum_{m=0}^{d} c_m P(2k, m)
\end{equation}
where again the matrix $P(n,m)$ is given by equation \eqref{viy}. We then substitute the reconstruction \eqref{ibt} to the expansion \eqref{bxi} and compute the finite part integrals \eqref{tuy} occurring in the second term of equation \eqref{bxi}, 
\begin{align}
    \sum_{k=0}^{\infty}\frac{(-1)^{k}\mu_{-(2k+2)}}{\beta^{k}} = &\sum_{k=0}^{\left \lfloor\frac{d}{2}\right\rfloor-1} \frac{(-1)^{k}}{\beta^{k}} \left(A_{2k} + B_{2k} + C_{2k} \right) + \sum_{k=\left \lfloor\frac{d}{2}\right\rfloor}^{\infty} \frac{(-1)^{k}}{\beta^{k}} D_{2k},
\end{align}
where $\lfloor x \rfloor$ is the floor function and the terms $A_k, B_k, C_k,$ and $D_k$ are given by equations \eqref{ah}-\eqref{dah}. 

We tabulated the result of using the expansion \eqref{bxi} to compute the ground-state energy in the strong-coupling regime in Table \ref{bangag1}. We included the results of the Pad\'e approximant $P^{25}_{26}(\beta)$ constructed from the PT series \eqref{tip} as well as the strong-coupling expansion \eqref{bami} from \cite{weniger1996construction} for comparison. Our result exhibits an excellent rate of convergence for coupling orders $\sim 10^{-2}-10^{-1}$ where even higher-order Pad\'e approximants and the strong-coupling expansion \eqref{bami} becomes unreliable. In the strong-coupling regime, our result successfully reproduces the first few digits of the result obtained from \cite{weniger1996construction}.

\begin{table}
\begin{tabular}{  c l }
		\hline
		Moments & $\beta = 0.01 $ \\ 
		\hline
		10 &  \textcolor{blue}{1.01 674 1}17  \\
		
		100 & \textcolor{blue}{1.01 674 136 375 472} 059  \\
		
		200 & \textcolor{blue}{1.01 674 136 375 473 20}2 826   \\
  
		500 & \textcolor{blue}{1.01 674 136 375 473 203 167 183} 427  \\
		
		1000 & \textcolor{blue}{1.01 674 136 375 473 203 167 181 798 1}46 029 \\
		
		1500 & \textcolor{blue}{1.01 674 136 375 473 203 167 181 798 150 975} 360  \\
		
		2000 & \textcolor{blue}{1.01 674 136 375 473 203 167 181 798 150 975 91}2 655  \\
		
		2500 & \textcolor{blue}{1.01 674 136 375 473 203 167 181 798 150 975 913 335} 65  \\
        \hline

        $\mathcal{E}(\beta)$ & 0.821 842 620 537 508  \\
		\hline
		$P^{25}_{26}(\beta)$ & 1.01 674 136 331 858 137 175 447 611 \\
        \hline
	\end{tabular}

\begin{tabular}{  c l l }

		Moments & $\beta = 0.1$ & $\beta = 1.0$\\ 
		\hline
		10  & \textcolor{blue}{1.10} 893 & \textcolor{blue}{1.43} 049 \\
		
		100   & \textcolor{blue}{1.10 908 70}5 248 &  \textcolor{blue}{1.43 5}58 957\\
		
		200 & \textcolor{blue}{1.10 908 707 8}08 &  \textcolor{blue}{1.43 562 470}  \\

		500 & \textcolor{blue}{1.10 908 707 846} 624 &  \textcolor{blue}{1.43 562 470} \\
		
		1000 & \textcolor{blue}{1.10 908 707 846 558 }285  &  \textcolor{blue}{1.43 562 461} 738\\

		1500  & \textcolor{blue}{1.10 908 707 846 558 3}69 &  \textcolor{blue}{1.43 562 461} 895\\
		
		2000  & \textcolor{blue}{1.10 908 707 846 558 370 2}53 &  \textcolor{blue}{1.43 562 461} 898 \\
		
		2500 & \textcolor{blue}{1.10 908 707 846 558 370 28}0 768 &\textcolor{blue}{1.43 562 461 900} \\
        \hline

        $\mathcal{E}(\beta)$ & 1.10 907 981 161 923  & 1.43 562 461 87 \\
		\hline
		$P^{25}_{26}(\beta)$ &  1.10 864 422 246 116 & 1.33 992 038 209  \\
        \end{tabular}

 	\begin{tabular}{  c  l l l l l }
        \hline
		Moments  & $\beta = 100$ & $\beta = 10^{3}$ & $\beta = 10^{4}$ & $\beta = 10^{5}$ & $\beta = 10^{6}$  \\ 
		\hline
		10   & \textcolor{blue}{3.}58541 & \textcolor{blue}{6.}15915 & \textcolor{blue}{1}0.7643 & 18.9688 & \textcolor{blue}{3}3.5672\\

		100 &  \textcolor{blue}{3.7}0518 & \textcolor{blue}{6.4}4745 & \textcolor{blue}{11.}3605 & \textcolor{blue}{20.}1160 & \textcolor{blue}{3}5.6959 \\
		
		200  & \textcolor{blue}{3.71}338 & \textcolor{blue}{6.4}7582 & \textcolor{blue}{11.4}310 & \textcolor{blue}{20.}2654 & \textcolor{blue}{3}5.9878  \\

		500  & \textcolor{blue}{3.71}758 &  \textcolor{blue}{6.49}604 & \textcolor{blue}{11.4}899 & \textcolor{blue}{20}.4025 & \textcolor{blue}{36.2}722 \\

		1000  & \textcolor{blue}{3.7169}0 & \textcolor{blue}{6.49}176 & \textcolor{blue}{11.47}65 & \textcolor{blue}{20.3}693 & \textcolor{blue}{36.}1989   \\

        1500  & \textcolor{blue}{3.71697}123 & \textcolor{blue}{6.4923}2 & \textcolor{blue}{11.4787}086 & 
         \textcolor{blue}{20.37}48  & \textcolor{blue}{36.211}1 \\
        
		\hline
		$\mathcal{E}(\beta)$& 3.71697473 &  6.49235 & 11.4787980  & 20.3750   &  36.2115 \\
		\hline
	    $P^{25}_{26}(\beta)$  & 1.46684099 & 1.46847 & 1.46863451 & 1.46865  & 1.46865  \\
		\hline
	\end{tabular}
    	\begin{tabular}{ cllllc }
	
		Moments  & $\beta = 10^{8}$  & $\beta = 10^{9}$ & $\beta = 10^{10} $ & $\beta = 10^{11}$ & $\beta=2(10^{20})$  \\ 
		\hline
		50 & \textcolor{blue}{11}6.073 & \textcolor{blue}{20}6.457 & \textcolor{blue}{36}7.188  & \textcolor{blue}{6}53.014 & \textcolor{blue}{13}8110  \\ 		
		100 &  \textcolor{blue}{11}2.687  & \textcolor{blue}{20}0.432 & \textcolor{blue}{3}56.167  & \textcolor{blue}{6}33.30 & \textcolor{blue}{13}3909 \\ 
		
		200 &  \textcolor{blue}{11}3.689 & \textcolor{blue}{20}2.132 & \textcolor{blue}{3}59.412 & \textcolor{blue}{6}39.10 &\textcolor{blue}{13}5144  \\

		300 &  \textcolor{blue}{114.}854 & \textcolor{blue}{20}4.255 & \textcolor{blue}{36}3.237  & \textcolor{blue}{64}5.95& \textcolor{blue}{136}607\\ 
		
		500 & \textcolor{blue}{114.}696 & \textcolor{blue}{20}4.025 & \textcolor{blue}{362}.723  & \textcolor{blue}{64}5.03& \textcolor{blue}{136}411 \\ 
	
		1000 &  \textcolor{blue}{114.4}36 & \textcolor{blue}{203.}969 & \textcolor{blue}{36}1.861  & \textcolor{blue}{643.}49 & \textcolor{blue}{136}080\\
		
		1500 &  \textcolor{blue}{114.48}2 & \textcolor{blue}{203.57}6 & \textcolor{blue}{362.01}4 & \textcolor{blue}{643.7}6 & \textcolor{blue}{1361}39 \\
		\hline
		$\mathcal{E}(\beta)$ & 114.483  & 203.579 & 362.019  & 643.77 & 136141  \\
		\hline
		$P^{25}_{26} (\beta)$ & 1.46865 & 1.46865 & 1.46865 & 1.46865 &  1.46865\\
		\hline
	\end{tabular}

	\caption{Convergence of the expansion \eqref{bxi} by finite-part integration of the ground-state energy of the sextic anharmonic oscillator across all coupling regimes. All relevant computations are performed up to 3000-digit precision. We obtained $\mathcal{E}(\beta)$ from the result of \cite{weniger1996construction}. $P_{26}^{25}(\beta)$ is a Pad\'e approximant.}
	\label{bangag1}
\end{table}
\section{Summary, Recommendations, and Conclusion}\label{summ}
In this paper we proposed a prescription for applying the method of finite-part integration by summing divergent Stieltjes series to a Stieltjes integral or one of its generalizations. We applied the procedure on divergent PT series solutions for the ground-state energy of various quantum mechanical systems. The procedure allowed us to transform the divergent weak-coupling expansion into a novel convergent strong-coupling expansion in inverse powers of the coupling parameter plus a correction term that led us to naturally incorporate the known non-integer algebraic leading-order behavior of the energy eigenvalues in the strong-coupling regime. We then computed approximations to the ground-state energies across all perturbation regimes with considerable accuracy even well into the strong-coupling regime. 

We pointed out how the point-wise convergence of the solution to the underlying Stieltjes moment problem near the origin could undermine the convergence of the expansion in the strong-coupling regime. In this regard, future work on the subject could investigate procedures for solving the Stieltjes moment problem that could offer better point-wise convergence with less number of positive-power moments used as inputs. This could enhance the potential of the prescription to take on problems in certain applications where the perturbation coefficients are scarce. In addition, the reconstruction must possess the required analytic property for finite part integration to apply and the appropriate form at the origin to guarantee that it can simulate the correct strong-coupling leading-order behavior of the problem. The maximum entropy approach \cite{mead1984maximum} for instance, gives the reconstruction in the form
\begin{equation}\label{hoppt}
    \rho(x) = \exp \left( -\sum_{n=0}^{N}{\lambda_n} x^{n } \right),
\end{equation}
where $\lambda_n$ are constants to be determined. This is not an ideal reconstruction scheme when used for the physical examples considered here. When $\rho(x)$ is given by \eqref{hoppt}, finite part integration gives the expansion for the Stieltjes integral as (see \cite{galapon2}, eq. 3.3),
\begin{equation}
 \beta\int_{0}^{\infty}\frac{\rho(x)}{1+\beta x}\mathrm{d}x =\sum_{k=0}^{\infty}\frac{(-1)^{k}}{\beta^{k}} \mu_{-(k+1)} - \rho\left(-\frac{1}{\beta}\right)\ln\left(\frac{1}{\beta}\right).
\end{equation}
Where the negative-power moments are the Hadamard's finite part integral (eq. 2.9 in \cite{galapon2})
\begin{equation}
    \mu_{-(k+1)} = \bbint{0}{\infty} \frac{\rho(x)}{x^{k+1}}\mathrm{d}x.
\end{equation}
In the strong-coupling regime, this expansion will be dominated by the second term which exhibits a logarithmic behavior instead of the non-integer algebraic power exhibited in the spectra of the systems considered here. 

The versatility and the comparative algorithmic simplicity gives the method of summation we presented here its obvious advantage. Through the most direct route, it allowed us to bring partial information from opposite asymptotic regimes and yield a single expansion which can be used to compute the solution across all parametric regimes. Moreover, as shown for the case of the spectrum of the sextic anharmonic oscillator, the prescription can also be applied efficiently to sum series which exhibits divergence faster than a factorial growth for which standard Pad\'e approximants could not handle effectively even in the small-coupling regime. 

\section*{Acknowledgments}
We acknowledge the Computing and Archiving Research Environment (COARE) of the Department of Science and Technology's Advanced Science and Technology Institute (DOST-ASTI) for providing access to their High-Performance Computing (HPC) facility. This work was funded by the University of the Philippines System through the Enhanced Creative Work Research Grant (ECWRG 2019-05-R). C.D. Tica acknowledges the Department of Science and Technology-Science Education Institute (DOST-SEI) for the scholarship grant under DOST ASTHRDP-NSC. 
\appendix

\section{Hadamard's Finite Part Integrals}\label{hfp}

In this section, we give a concise discussion on the computation of Hadamard's finite part of the divergent integrals with a non-integrable singularity at the origin,
\begin{equation}\label{miv}
    \int_{0}^{a} \frac{f(x)}{x^{m+\nu}} \mathrm{d}x,\qquad m=1,2,\dots,\, |\nu|<1,
\end{equation}
where $a$ can be taken to infinity. This is done by introducing an arbitrarily small cut-off parameter $\epsilon$, $0<\epsilon<a$, to replace the offending non-integrable origin. The resulting convergent integral is grouped into two sets of terms
\begin{equation}\label{definitepart}
\int_{\epsilon}^{a} \frac{f(x)}{x^{m+\nu}} \mathrm{d}x = C_{\epsilon}+D_{\epsilon},
\end{equation}
where $C_{\epsilon}$ is the group of terms that possesses a finite limit as $\epsilon\rightarrow 0$, while $D_{\epsilon}$ diverges in the same limit. The finite part of the divergent integral is then obtained by dropping the diverging group of terms $D_{\epsilon}$, leaving only the limit of $C_{\epsilon}$ and assigning the limit as the value of the divergent integral,
\begin{align}\label{finitepart}
\bbint{0}{a} \frac{f(x)}{x^{m+\nu}}\mathrm{d} x = \lim_{\epsilon\rightarrow 0} C_{\epsilon} .
\end{align}
The Hadamard's finite part integral may also be represented more rigorously as the complex contour integral given in equation \eqref{result2} in Section \ref{meth}. The details are given in \cite{galapon2, galapon2016cauchy}. The starting point for the construction of a complex contour integral representation is the equivalent form,
\begin{align}\label{form2}
\bbint{0}{a} \frac{f(x)}{x^{m+\nu}} \mathrm{d}x = \lim_{\epsilon\rightarrow 0} \left[\int_{\epsilon}^a \frac{f(x)}{x^{m+\nu}} \mathrm{d}x - D_{\epsilon}\right].
\end{align}
The canonical representation \eqref{finitepart}, on the other hand is relevant for computing the values of the Hadamard's finite part explicitly. To demonstrate this, let  $f(x)$ have an entire complex extension $f(z)$ such that it admits the expansion 
\begin{equation}
f(z)=\sum_{k=0}^{\infty} c_k z^k .
\end{equation}
Substituting this expansion in equation\eqref{miv}, we see from equation \eqref{finitepart} that
\begin{equation}\label{branchpoint}
\bbint{0}{a} \frac{f(x)}{x^{m+\nu}} \, \mathrm{d}x = \sum_{k=0}^{\infty} \frac{c_k a^{k+1-m-\nu}}{(k+1-m-\nu)}.
\end{equation}
Moreover,
\begin{align}\label{infinity_si_a}
\bbint{0}{\infty} \frac{f(x)}{x^{m+\nu}} \, \mathrm{d}x = \lim_{a\rightarrow\infty} \sum_{k=m}^{\infty} \frac{c_k a^{k+1-m-\nu}}{(k+1-m-\nu)},
\end{align}
provided $f(x) x^{-m-\nu}$ is integrable at infinity or $f(x) x^{-m-\nu}=o(x^{-1})$ as $x\rightarrow\infty$.
We apply this to the computation of the finite part for which $f(x) = e^{-b\,x}$, $b>0$. Substituting the coefficients $c_k=(-b)^k/k!$ back in equation \eqref{infinity_si_a}, we obtain the finite part in limit form
\begin{align}\label{motyin}
	\bbint{0}{\infty}\,\frac{e^{-b\,x}}{x^{m+\nu}}\,\mathrm{d}x = \lim_{a\to\infty}\,\sum_{k=m}^{\infty}\,\frac{\left(-1\right)^{k}\,b^{k}\,a^{k+1-m-\nu}}{k!\,\left(k+1-m-\nu\right)} .
\end{align}
To facilitate the calculation of the limit, we sum the series as a hypergeometric function,
\begin{equation}\label{vivy}
\sum_{k=m}^{\infty}\,\frac{\left(-1\right)^{k}\,b^{k}\,a^{k+1-m-\nu}}{k!\,\left(k+1-m-\nu\right)} = \frac{\left(-1\right)^{m+1}\,a^{1-\nu}\,b^{m}}{m!\,\left(\nu-1\right)} \, \pFq{2}{2}{1,1-\nu}{m+1, 2-\nu}{-a b},
\end{equation}
and make use of its asymptotic expansion for large arguments for the case of simple poles in \cite{simplepole2F2},
 \begin{eqnarray}	\pFq{2}{2}{a_1,a_2}{b_1, b_2}{z}&=& \frac{\Gamma\left(b_1\right)\,\Gamma\left(b_2\right)\,\Gamma\left(a_2-a_1\right)}{\Gamma\left(a_2\right)\,\Gamma\left(b_1-a_1\right)\,\Gamma\left(b_2-a_1\right)}\,\left(-z\right)^{-a_1}\,\left(1+\mathcal{O}\left(z^{-1}\right)\right)\nonumber\\
&&    +\,\,\frac{\Gamma\left(b_1\right)\,\Gamma\left(b_2\right)\,\Gamma\left(a_1-a_2\right)}{\Gamma\left(a_1\right)\,\Gamma\left(b_1-a_2\right)\,\Gamma\left(b_2-a_2\right)}\,\left(-z\right)^{-a_2}\,\left(1+\mathcal{O}\left(z^{-1}\right)\right)\nonumber\\
&&    +\,\,\frac{\Gamma\left(b_1\right)\,\Gamma\left(b_2\right)}{\Gamma\left(a_1\right)\,\Gamma\left(a_2\right)}\,e^{z}\,z^{a_1+a_2-b_1-b_2}\,\left(1+\mathcal{O}\left(z^{-1}\right)\right)
\end{eqnarray}
By inspection, the second term in this expansion dominates all the others for the given parameters in equation \eqref{vivy}. Hence in the limit as $a\rightarrow\infty$,
 \begin{align}
     \pFq{2}{2}{1,1-\nu}{m+1, 2-\nu}{-a b} = \frac{\Gamma(m+1) \Gamma(2-\nu)\Gamma(\nu)}{\Gamma(m+\nu)\,(ab)^{1-\nu}}
 \end{align}
 so that we obtain the finite part integral \eqref{motyin}
\begin{align}\label{expofinite}
	\bbint{0}{\infty}\,\frac{e^{-bx}}{x^{m+\nu}}\,\mathrm{d}x = \frac{\left(-1\right)^{m}\,b^{m+\nu-1}\,\pi}{\sin\left(\pi\,\nu\right)\,\Gamma\left(m+\nu\right)}, \qquad b>0,
\end{align}
$m=1,2,\dots$ and $|\nu|<1.$

Alternatively and more conveniently, the finite part \eqref{expofinite} can also be extracted from the analytic continuation of the Mellin transform \cite{brychkov2018handbook},
\begin{eqnarray}\label{monir}
    \mathcal{M}\left[e^{-bx};1-\lambda\right]&=&\int_0^{\infty} x^{-\lambda} e^{-b x}\,\mathrm{d}x\nonumber\\
    &=&b^{\lambda-1} \Gamma(1-\lambda), \;\; \operatorname{Re}\lambda<1.
\end{eqnarray}
As established in \cite{regularizedlimit}, the finite part of the divergent integral $\int_0^{\infty} x^{-\lambda} e^{-b x}\,\mathrm{d}x$ is the analytic continuation of the right-hand side of  equation 
 \eqref{monir} to $\operatorname{Re}\lambda \geq 1$ provide $\lambda$ is a regular point of the analytic continuation. The result \eqref{monir} reduces to \eqref{expofinite} by making use of the identity
 \begin{align}
     \Gamma(1-\lambda) = \frac{\pi}{\sin(\pi\lambda)\Gamma(\lambda)}
 \end{align}
 and setting $\lambda = m + \nu$. In this particular case, the latter method is obviously more convenient for the computation of the finite part integral \eqref{expofinite}. But for cases when the Mellin transform does not exist, the canonical form \eqref{branchpoint} is a convenient alternative.


\end{document}